\documentclass[12pt,preprint]{aastex}

\usepackage{graphicx}

\newcommand{\dxdy}[2]{{\frac{\partial{#1}}{\partial{#2}}}}
\newcommand{\dxdys}[2]{{\frac{\partial^{2}{#1}}{\partial^{2}{#2}}}}
\newcommand{\DxDy}

\shorttitle{Angular Momentum transport by Internal Gravity waves in the Sun}
\shortauthors{Rogers et al.}

\begin{document}

\title{Angular Momentum Transport by Gravity waves in the Solar Interior}

\author{Tamara M. Rogers} 
\affil{Astronomy and Astrophysics Department, University of California,
    Santa Cruz, CA 95064}
\email{trogers@pmc.ucsc.edu}
\author{Gary A. Glatzmaier}
\affil{Earth Sciences Department, University of California, Santa Cruz, 
	CA 95064}

\begin{abstract}

We present self-consistent numerical simulations of the sun's convection zone and radiative interior using a two-dimensional model of its equatorial plane.   The background reference state is a one-dimensional solar structure model.  Turbulent convection in the outer convection zone continually excites gravity waves which propagate throughout the stable radiative interior and deposit their angular momentum.  We find that angular velocity variations in the tachocline are driven by angular momentum transported by overshooting convective plumes rather than the nonlinear interaction of waves.  The mean flow in the tachocline is time dependent but not oscillatory in direction.  Since the forcing in this shallow region can not be described by simple linear waves, it is unlikely that the interaction of such waves is responsible for the solar cycle or the 1.3 year oscillation.  However, in the deep radiative interior, the interaction of low amplitude gravity waves, continually excited by the overshooting plumes, is responsible for the angular velocity deviations observed there.  Near the center of the model sun the angular velocity deviation is about two orders of magnitude greater than that in the bulk of the radiative region and reverses its direction (prograde to retrograde or vice versa) in the opposite sense of the angular velocity deviations that occur in the tachocline.  Our simulations thus demonstrate how angular velocity variations in the solar core are linked to those in the tachocline, which themselves are driven by convective overshooting.  

\end{abstract}


\section{Introduction}

Internal gravity waves are ubiquitous in nature.  Their influence can be observed in striated cloud structures in our own atmosphere many days of the year.  In addition to the visual display of these waves in our atmosphere, they have several other more profound consequences.  The interaction of gravity waves with angular velocity shear produces the quasi-biennial oscillation (QBO), which dominates variability in the equatorial stratosphere and affects ozone levels.  

Gravity waves have been invoked to explain several physical problems in the sun's interior such as: (1) providing the extra mixing required to solve the Li depletion problem (Garcia-Lopez \& Spruit 1991) (2) increasing turbulent mixing and affecting the solar neutrino production (Press 1981), (3) maintaining the solid body rotation of the sun's radiative interior (Schatzman 1993, Kumar \& Quataert 1997) (4) and controlling the solar cycle (Kumar, Talon \& Zahn 1999).  They have also been used to explain the orbital properties of binary star systems (Terquem et al. 1998).  These theories, however, are only as good as the poorly understood excitation and evolution of gravity waves in stellar interiors.  What gravity wave spectra and amplitudes are generated?  How effectively do gravity waves transport angular momentum and mix species?

Some confusion has existed.  For example, although gravity waves were postulated to extract angular momentum from the solar radiative interior and so enforce solid body rotation (Schatzman 1993, Kumar \& Quataert 1997), it was quickly pointed out (Gough \& McIntyre 1998, Ringot 1998) that gravity waves tend to enhance local shear rather than smooth it.

Recognizing this anti-diffusive nature of gravity waves in shear flows, several authors published papers postulating an angular velocity oscillation at the base of the solar convection zone analogous to the quasi-biennial oscillation (QBO) in the Earth's stratosphere (Baldwin et al. 2001).  In one paper (Kumar, Talon \& Zahn 1999), a spectrum of gravity waves generated by the overlying convection is prescribed (Goldreich et al. 1994).  This spectrum is then integrated to give a flux of angular momentum that is transferred from the waves to the mean flow, resulting in a periodic oscillation of angular velocity at the base of the convection zone with a timescale of about 20 years.  In another paper (Kim \& MacGregor 2001), a two wave model is assumed: one prograde propagating wave, with a prescribed angular momentum flux, and one retrograde wave with another (negative) prescribed flux.  In addition, these authors include viscous dissipation in the evolution equation for the mean flow and find that the nature of the resulting angular velocity oscillation depends sensitively on the value of the assumed viscous diffusivity.  For large viscous diffusivity, a steady solution is found; whereas for small values a chaotic oscillation is found.  A periodic solution is recovered only for intermediate values of the viscous diffusivity.  

More recently, the Kumar, Talon \& Zahn (1999) model has been extended to show that the oscillating shear layer at the base of the convection zone acts as a filter on the low frequency, short wavelength waves (Talon, Kumar \& Zahn 2002, Talon \& Charbonnel 2005).  In their theory, the filter preferentially damps prograde waves, allowing predominantly retrograde waves, which carry negative angular momentum, into the deep interior.  When these waves dissipate they transfer their negative angular momentum to the flow and therefore effectively extract prograde angular momentum from the low-latitude deep solar interior, leading to solid body rotation in the solar radiative zone.  

The major shortcoming of both of these models is twofold.  First, neither model self-consistently calculates the generation of the gravity waves by the solar convection zone.  In Kumar, Talon \& Zahn (1999) a spectrum and amplitude of gravity waves is assumed, a spectrum that is unfortunately, untestable with observations and has not been reproduced in numerical simulations.  In addition, this spectrum neglects the main production of gravity waves due to overshooting plumes.  Second, and perhaps more importantly, neither model calculates the nonlinear wave-wave interactions which provide this ``flux'' of angular momentum.  Rather than parameterizing the nonlinear interaction of waves as a prescribed flux (Kumar, Talon \& Zahn 1999 and subsequent papers, Kim \& MacGregor 2001), this interaction should be self-consistently calculated.

Here we present numerical simulations that address both of these issues.  Our model solves the fully nonlinear Navier-Stokes equations in both the convective and radiative regions.  Therefore, the gravity waves are self-consistently generated by an overlying convection zone and the nonlinear terms are retained in the radiation zone to account for the nonlinear wave interactions which affect the mean flow.  

\section{Wave - shear flow interactions}

In this section we briefly review the hydrodynamic process by which waves can transport their angular momentum to the flow in which they travel.  The natural example of this process is the QBO.  For a more complete discussion of the QBO see, for example, Baldwin et al. 2001, Lindzen (1990) and Holton (1994).  

For simplicity consider the Boussinesq equations for a viscous fluid in two dimensional cartesian (x-z) coordinates, x being the horizontal coordinate and z the vertical coordinate.  After decomposing the horizontal velocity into mean flow ($\overline{U}$, a function of z) and fluctuating (u') components and taking the proper horizontal average, the equation for the mean horizontal velocity becomes:

\begin{equation}
\dxdy{\overline{U}}{t}=-\dxdy{\overline{u'w'}}{z}+\nu\dxdys{\overline{U}}{z}
\end{equation}

In (1), ${u'}$ represents the fluctuating horizontal velocity and ${w'}$ represents the fluctuating vertical velocity.  This equation represents how small scale wave-wave interactions (Reynolds stresses) influence the mean flow ($\overline{U}$).  The vertically propagating waves carry horizontal momentum vertically.  Eliassen \& Palm (1961) showed that (in this simplified model with no viscosity) the first term on the right hand side of (1) is zero unless there is some wave attenuation (such as radiative damping or critical layers).  Therefore, in the absence of wave attenuation, there is no transfer of angular momentum between the mean flow, $\overline{U}$, and the Reynolds stresses.  In the sun the proposed mechanism for wave attenuation is via radiative diffusion, while in the Earth the process is likely wave breaking. 

In the presence of differential rotation the picture becomes more complicated.  In a rotating fluid, waves generated at a particular frequency are doppler shifted away from that frequency according to the equation:
\begin{equation}
\omega(r)=\omega_{gen}+m(\Omega_{gen}-\Omega(r))
\end{equation}
where $m$ is the horizontal wave number, $\omega_{gen}$ is the frequency at which the wave is generated in the frame rotating at $\Omega_{gen}$ and $\omega(r)$ is the frequency measured relative to the local rotation rate, $\Omega(r)$.

Where $\Omega(r) > \Omega_{gen}$ prograde waves ($m > 0$) are shifted to lower frequencies and retrograde waves ($m < 0$) are shifted to higher frequencies.  Since radiative damping is strongly frequency dependent (damping length $\propto \omega^{4}$), prograde (lower frequency) waves are damped in a shorter distance than retrograde (higher frequency) waves.  This differential damping causes prograde waves to deposit their (positive) angular momentum closer to the generation site than retrograde waves deposit their (negative) angular momentum, thus leading to a shear layer.  As prograde (retrograde) waves continually deposit their positive (negative) angular momentum, the angular velocity amplitude and gradient continually increase.  This increased angular velocity causes prograde waves to be shifted to ever smaller frequencies which are damped even closer to the generation site.  In this way the peak in the prograde layer moves toward the source of the waves.  When the prograde shear becomes sufficiently steep, it is broken down by viscous diffusion, leaving behind the retrograde layer.  This process repeats, with the period of the process inversely proportional to the wave forcing.  The prominent features of this physical process are then twofold: (1) prograde flow lies above retrograde motion (or vice versa) and (2) this pattern propagates toward the generation site.  

This physical picture was proposed initially by Lindzen \& Holton (1968) and Holton \& Lindzen (1972) to explain the QBO observed in the Earth's atmosphere.  Later, the physical theory was tested in the remarkable experiment by Plumb \& McEwan 1978 and the basic physical mechanism was recovered.  It has also been suggested that an oscillation similar to the QBO occurs in Jupiter's atmosphere (Leovy et al. 1991) and is coined the quasi-quadrennial oscillation (QQO) because of its four year period.

The robustness of this mechanism led astronomers to hypothesize that the same physical mechanism could be operating in the solar tachocline (Kumar, Talon \& Zahn 1999, Kim \& MacGregor 2001, Talon \& Charbonnel 2005).  This could then provide a handsome explanation for oscillations at the base of the convection zone, whether on 20 year timescales (as in the dynamo, Kumar, Talon \& Zahn 1999) or 1 year timescales (as in the 1.3 year oscillation, Kim \& MacGregor 2001).  We have reviewed this mechanism here so comparisons between this process and the zonal flow oscillations seen in the radiative region of our model can be clearly made.   

\section{Numerical Model}

The numerical technique and model setup are identical to those in Rogers \& Glatzmaier 2005, except here we impose the equatorial rotation profile as a function of radius as inferred from helioseismology ($\Omega(r)$ is specified to be 465nHz in the convection zone, 435nHz in the stable region and the tachocline is fit to an error function).  We solve the Navier-Stokes equations with rotation in the anelastic approximation in 2D cylindrical geometry (r,$\phi$).  The curl of the momentum equation, i.e., vorticity equation, is:
 \begin{equation}
\dxdy{\omega}{t}+(\vec{v}\cdot\vec{\nabla})\omega=(2\Omega(r) + \omega)h_{\rho}v_{r}-2v_{r}\dxdy{\Omega(r)}{r}-\frac{\overline{g}}{\overline{T}r}\dxdy{T}{\theta}-\frac{1}{\overline{\rho}\overline{T}r}\dxdy{\overline{T}}{r}\dxdy{{\it{p}}}{\theta} +\overline{\nu}\nabla^{2}\omega
\end{equation}

The heat equation is: 
\begin{eqnarray}
\lefteqn{\dxdy{T}{t}+(\vec{v}\cdot\nabla){T}=-v_{r}(\dxdy{\overline{T}}{r}-(\gamma-1)\overline{T}h_{\rho})+(\gamma-1)Th_{\rho}v_{r}+}\nonumber \\
&  & \gamma\overline{\kappa}[\nabla^{2}T+(h_{\rho}+h_{\kappa})\dxdy{T}{r}]+\gamma\overline{\kappa}[\nabla^{2}\overline{T}+(h_{\rho}+h_{\kappa})\dxdy{\overline{T}}{r}] + \frac{\overline{Q}}{c_{v}}
\end{eqnarray}

In these equations, $\vec{v}$ is the velocity, with radial, $v_r$, and longitudinal, $v_\theta$, components.  The vorticity is $\vec{\omega}=\vec{\nabla}\times\vec{v}$ and is normal to the equatorial plane in this 2D geometry.  The functions $h_{\rho}=dln\overline{\rho}/dr$, $h_{\kappa}=dln\overline{\kappa}/dr$, $\overline{g}$ (gravity), $\overline{T}$ (temperature), $\overline{\rho}$ (density) and $\gamma$ (ratio of specific heats, $c_p/c_v$) are radially dependent and taken from the solar model.   T is the temperature perturbation and ${\it p}$ is the pressure perturbation, which like $\omega$, are functions of $r$, $\theta$ and time ($t$).

In our model we specify the thermal diffusivity as that given by the solar model, multiplied by a constant for numerical stability: 
\begin{equation}
\overline{\kappa}=kapmult*\frac{16\sigma \overline{T}^{3}}{3\overline{\rho}^{2}\overline{k}c_{p}}
\end{equation}
here {\it kapmult} is generally set to $10^{5}$, $\sigma$ is the Stefan-Boltzman constant and $\overline{k}$ is the opacity.  The viscous diffusivity, $\overline{\nu} ={\mu}/\overline{\rho}$, is set so that the dynamic viscosity ($\mu$) is constant.   

We calculate the pressure term in (3) using the longitudinal component of the momentum equation: 
\begin{equation}
\frac{1}{\overline{\rho}r}\dxdy{\it{p}}{\theta}=-\dxdy{v_{\theta}}{t}-(\vec{v}\cdot\vec{\nabla}\vec{v})_{\theta}+\overline{\nu}[(\nabla^{2}\vec{v})_{\theta} - \frac{h_\rho}{3r}\dxdy{v_{r}}{\theta}] .
\end{equation}
These equations are supplemented by the continuity equation in the anelastic approximation
\begin{equation}
\nabla \cdot \overline{\rho} \vec{v} = 0 
\end{equation}
which is satisfied by expressing $\overline{\rho} \vec{v}$ as the curl of a streamfunction.  The model extends from .001$R_{\odot}$ to 0.93$R_{\odot}$.  The subadiabaticity in the radiative region is given by the solar model and we specify the superadiabaticity in the convection zone to be $10^{-7}$.  
  
These equations are solved using a Fourier spectral transform method in the longitudinal ($\theta$) direction and a finite difference scheme on a non-uniform grid in the radial (r) direction.  Time advancing is done using the explicit Adams-Bashforth method for the nonlinear terms and an implicit Crank-Nicolson scheme for the linear terms.  The boundaries are impermeable and stress-free.  The inner boundary is isothermal and the outer boundary is held at a constant heat flux.  This code is parallelized using message passing interface (MPI) and the resolution is 2048 longitudinal zones x 1500 radial zones, with 620 radial zones dedicated to the radiative region.  In the region just below the convection zone the radial resolution is 170km.  This model was evolved for 1 simulated year, requiring nearly six million 5 second timesteps.

\section{Convection zone and Tachocline}
\subsection{Angular Velocity Variations}

Helioseismic observations indicate that, in the equatorial plane, the convection zone spins faster than the radiative interior.  The maintenance of differential rotation within the equatorial plane of the convection zone is a 3D process involving the transport of angular momentum in both radius and latitude and the Coriolis forces resulting from axisymmetric meridional circulation.  Since our 2D geometry captures only the transport in radius, we can not expect to achieve a realistic differential rotation profile with the 2D model.  Therefore, we impose the observed equatorial angular velocity as a function of radius in our model.  As mentioned above, the background angular velocity is set to a constant 465nHz in the convection zone and to a constant 435nHz in the radiation zone; we prescribe a smooth fit between these two values through the thin tachocline shear layer.  The resulting gravity wave spectrum and angular momentum transport is then investigated.  

The time series of angular velocity, $\Omega^{'}(r,t)$, relative to the prescribed background angular velocity, $\Omega(r)$, for this model is shown in Figure 1 over one simulated year.  As seen in this figure, angular velocity in the convection zone is initially prograde, relative to the prescribed $\Omega(r)$, in the lower part of the convection zone and retrograde motion in the upper part.   However, after the initial period, angular velocity becomes very time dependent and is dominantly retrograde in the lower part of the convection zone and prograde in the upper part.  This profile is expected, given the density stratification and rotation of the solar interior (Glatzmaier et al. 2005).  Intermittently prograde motion from the top of the convection zone will extend to the base of the convection zone and overshoot into the tachocline.  The angular velocity variations produced by these motions vary in amplitude between +/- 15nHz, slightly larger than the amplitudes of the observed 1.3 year oscillation (Howe et al. 2001).  Figure 1a clearly shows that the angular velocity of the tachocline mimics the behavior in the lower part of the convection zone; when the lower part is prograde, the tachocline is prograde and vice versa.  This indicates that the behavior of the tachocline is dictated by convection zone dynamics, rather than by gravity waves (see below). 

In the overshoot region fluid motions are strongly nonlinear (Rogers \& Glatzmaier 2005b).  Figure 2 shows the ratio of the horizontal fluid velocity to the horizontal phase speed (the Froude number) for a typical frequency and wavenumber (20$\mu$Hz, l=10); this provides a measure of the nonlinearity of waves.  For linear waves $u_{x}/c_{x} << 1$.  However, as is clearly seen in the figure this criterion does not hold just below the convection zone.   While only one ratio is shown as a function of radius, this can vary by an order of magnitude depending on the choice of frequency and wavenumber.\footnote{For larger (smaller) values of horizontal wave mode number this ratio is larger (smaller).  Similarly, for larger (smaller) frequencies this ratio is smaller (larger).} However, using reasonable values for frequency and wavenumber, the smallest value of this ratio just below the convection zone is 0.1, still far from meeting the linearity criterion above. Furthermore, it is clear in this figure that linearization may not be justified in the solar core either, as suggested by Press (1981).  

\subsection{Reynolds and Viscous Stresses}

The mean zonal flow is determined by a balance between viscous and Reynolds stresses (as seen in equation (1)).  In order to better understand the angular velocity profile in time (as shown in Figure 1) we examine the sources of that balance.  Figure 3 shows the horizontally averaged convergence of Reynolds stress (first term on the right hand side of (1), but now in cylindrical coordinates), the viscous stress (second term on the right hand side of (1), in cylindrical coordinates) and the sum of these (the time rate of change of the mean zonal flow $\overline{U}$) in both the convection zone and overshoot region over five days. 

In the model's convection zone, the Reynolds stresses are much larger than the viscous stresses.  The pattern of Reynolds stress convergence has a semi-periodic behavior with disturbances propagating up and down from a radius near 1/3$d_{cz}$ ($d_{cz}$ is the convection zone depth).  When positive (i.e., prograde) Reynolds stress convergence moves upward from this radius a negative value moves downward and vice versa.  This process reverses with a frequency similar to the convective turnover frequency, roughly 10 times the inertial wave frequency.  This oscillatory behavior is observed in our simulations only when we include a radiative region beneath the convection zone; we do not see it when we impose an impermeable lower boundary on the convection zone.  Note also that this radius of 1/3$d_{cz}$ is approximately where $v_{\theta}$ for the large eddies changes sign and below which smaller, counter-rotating cells exist.  Figure 4 shows a zoomed-in region of the convection zone, displaying the vorticity within a few convective cells.  Beneath the large cells spanning most of the convection zone is counter-rotating fluid that is driven by the plumes penetrating slightly into the stable overshoot region.  As a plume rebounds it is diverted laterally and counter-rotating eddies are generated.  It remains to be seen how these counter-rotating cells affect tachocline dynamics in a 3D simulation.

Now consider the overshoot region in our model.  Regions with positive (negative) Reynolds stress convergence in the lower convection zone correspond to the same signed disturbances in the overshoot region.  These disturbances have a slight tendency to propagate downward, i.e., away from their source in the convection zone.  When viscous terms are added (which in the overshoot region have magnitudes similar to the Reynolds stress terms) disturbances clearly propagate away from the convection zone (Fig. 3).  This motion away from the convection zone \footnote{However, even if diffusion were neglected the disturbances do not move upward as in the QBO.} is also seen in Figure 1.  
The semi-periodic oscillations observed in $\partial{\overline{U}}/\partial{t}$ affects only the amplitude of $\overline{U}$, but not the direction.  That is, the semi-periodic oscillation seen $\partial{\overline{U}}/\partial{t}$ is not realized in $\overline{U}$ because of a slight asymmetry in the driving amplitude of prograde and retrograde motions.  If, as has been assumed in the past (Kumar, Talon \& Zahn 1999, Kim \& MacGregor), these motions were produced with equivalent amplitudes, the semi-periodic oscillation observed in $\partial{\overline{U}}/\partial{t}$ would be reproduced in $\overline{U}$, however, that is not the case here.  

Given the above results it is clear that overshooting plumes themselves transport angular momentum in a way which is distinct from the nonlinear interaction of low amplitude waves.  The main properties of a wave driven oscillation (motion in time toward the source, timescales different than the forcing timescales, and a double peaked layer) are not seen here.  Therefore, given the link between Reynolds stresses in the convection zone and those in the tachocline, it is clear that the overshooting plumes provide the main transfer of angular momentum to the tachocline.  These plumes are buoyantly braked, radiatively damped and the nonlinearity allows for mode-mode transfer which affects the damping rate.  Furthermore, the momentum transport by these plumes is time dependent and complex.  Their effects need to be included when considering the dynamics of the tachocline and their influence should not be treated simply as an increased flux in linear waves.  In fact, the linear approximation of waves is not justified in the overshoot region and tachocline.

\section{The Deep Interior and Core}

A banded radial differential rotation profile is observed deep within the radiative interior (Figure 5); however, the amplitude of the angular velocity variation is extremely low (0.02 nHz) relative to what it is in the convection zone.  This radial shear grows in time (Figure 6) due to wave-wave interactions which enforce shear.  However, there is some indication that the growth is slowing and possibly reversing (Figure 6) due to increased viscous diffusion as angular velocity gradients increase.  The banded differential rotation is due to the broad spectrum of frequencies generated and their {$\it continual$} deposition of angular momentum due to radiative diffusion as they propagate.  Banded differential rotation in the sun's radiative interior has not been inferred from helioseismology, possibly because the amplitude is too low, or because these small amplitude disturbances are easily neutralized in 3D.  In the core of our model the amplitude of the angular velocity variation is 2-3 orders of magnitude higher than in the bulk of the radiative interior.  This indicates that angular momentum transport from waves to zonal flow is more efficient in the core, which could be due to a number of factors, including: wave breaking, critical layers (where the phase speed of the wave approaches the mean rotation rate) or inefficient reflection as the wave frequency approaches the Brunt-Vaisala frequency.  Discerning which of these processes is dominant in this region is extremely difficult when a large spectrum of waves is continually being excited and is beyond the scope of this paper.  

Maxima in the convergence of Reynolds stress deep within the radiative interior (but a significant distance below the overshoot region) propagate upward (Figure 7) as expected from the nonlinear interaction of low amplitude waves.   In addition, significant interference is seen in the lower radiation zone, which likely comes from the interaction of inward propagating and outward reflected waves.  While we have not observed a complete reversal of the angular velocity in the bulk of the radiative region, there is some evidence that one is underway (Figure 6)\footnote{It appears that in the deep interior a QBO-like oscillation is possible}.  In the core the mean angular velocity switches from retrograde to prograde at nearly the same time as the tachocline switches from prograde to retrograde.  This suggests that the angular velocity of the core is linked to the angular velocity of the tachocline and hence, of the convection zone.  The link between tachocline angular velocity and that of the core due to selective filtering has been elucidated previously in Talon, Kumar \& Zahn 2002.  Here we make the link between the tachocline and the convection zone and propose that a self-consistent study of differential rotation in the sun must not treat the radiative and convective regions separately.

\section{Discussion}

Despite some similarity between the convective-radiative interface in the sun and the Earth's tropopause, there are several obvious differences.  In the sun, convection is constantly driving gravity waves everywhere below the overshoot region; whereas the generation of large-amplitude gravity waves in the Earth's atmosphere is intermittent in time and space.  Solar gravity waves travel down into a converging region of increasing density (and therefore, have very little chance of breaking); whereas in the Earth waves travel up into an expanding region of decreasing density.  Furthermore, previous numerical simulations of convective penetration in a stratified atmosphere demonstrate a remarkable difference between penetration into an overlying stable region and penetration into an underlying stable region (Hurlburt, Toomre \& Massaguer 1986).  Penetration into an underlying stable region is characterized by thin localized downflows; whereas penetration into an overlying stable region is characterized by larger scale broad upflows.  This asymmetry allows descending plumes to travel farther into an underlying stable region than ascending motions travel into an overlying stable region.  Therefore, it is likely that penetrative convection plays a more crucial role at the base of the solar convection zone than it does at the Earth's tropopause.  These differences can have profound effects on the role of overshoot and, hence, on the scale, frequency and amplitude of the waves generated and, so, on the angular momentum transport by these waves.  

Numerical simulations (Wedi \& Smolarkiewicz 2005) of the Plumb-McEwan laboratory experiment attempting to reproduce the QBO, have shown that the type and period of an oscillation in the differential rotation profile depend sensitively on the forcing.  In particular, it is found that random forcing rarely produces a periodic oscillation.  Given the turbulent nature of the sun, it is likely that the forcing is fairly random.  

For reasons stated above, it is unlikely that there is a QBO-like oscillation associated with the solar tachcocline.  However, QBO-like oscillations may occur in stars with radiative envelopes because of the inefficiency of overshoot into an overlying stable region, and because of more similar geometry.

\section{Conclusions}

We have presented self-consistent numerical simulations of convective overshoot and gravity wave generation and the angular momentum transport by these processes in a 2D model of the dynamics in the solar equatorial plane.  We find that angular velocity variations in the tachocline are driven by angular momentum transported by overshooting plumes rather than nonlinear interaction of low amplitude waves.  These overshooting plumes are strongly nonlinear disturbances, which can not be accurately represented as an increased flux of linear waves.   We observe a semi-periodic oscillation in amplitude, but not in direction, of the mean flow in the tachocline because of an asymmetry in the driving of prograde and retrograde motions.  Since we find that linear gravity waves are not dominant in the tachocline it is unlikely that they are responsible for the 1.3 year oscillation or the 11 year solar cycle.  It is no surprise that overshooting motions play a dominant role in the tachocline and we expect these results will persist in three-dimensions.  

In the deep radiative interior the continual deposition of angular momentum by the nonlinear interaction of gravity waves produces a radially banded differential rotation.  However, it remains to be seen whether this pattern persists in 3D, considering it's very low amplitude.  In the model's core, the amplitude of the differential rotation (i.e., angular velocity) is larger, about two orders of magnitude larger than that in the bulk of the radiative region and similar to the magnitude within the convection zone.  We observe retrograde motion in the core reversing to prograde motion in step with the counter-reversal (prograde to retrograde) at the tachocline.  When there is predominantly prograde flow at the base of the convection zone it selectively filters out prograde propagating gravity waves, allowing predominantly retrograde waves to propagate to the core where they deposit their (negative) angular momentum, and vice versa (Talon, Kumar \& Zahn 2002).  Like previous results our simulations therefore suggest that the angular velocity variations in the solar core are linked to those in the tachocline.  However, unlike previous results we show here that the variations in the tachocline are driven by convective overshooting, therefore, linking core rotation to convective motions.
  
\acknowledgments
We thank K. MacGregor, J. Christensen-Dalsgaard, D. Gough, C. Jones, P. Garaud and M. Metchnik for helpful discussions and guidance.  T.R. would like to thank the NPSC for a graduate student fellowship and the Institute of Astronomy, Cambridge University for a travel grant.  Support has also been provided by the DOE SciDAC program DE-FC02-01ER41176, the NASA Solar and Heliospheric Program SHP04-0022-0123  and the UCSC Institute of Geophysics and Planetary Physics.  Computing resources were provided by NAS at NASA Ames and by the NSF MRI grant AST-0079757.

\clearpage
\begin{figure}
\includegraphics[width=5in]{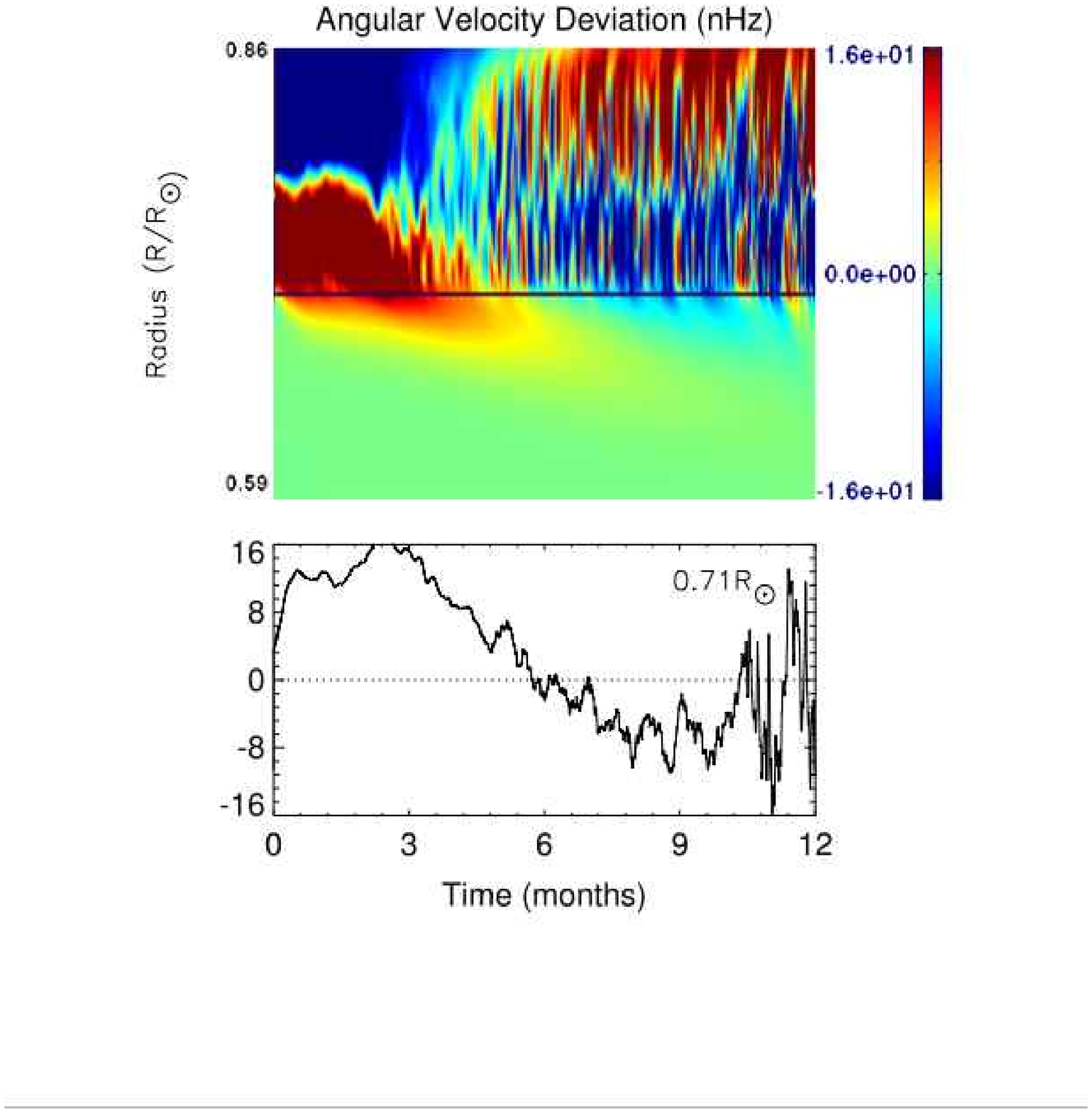}
%\plotone{figure1.jpg}
\caption{Top: Angular velocity variations, relative to the prescribed solar profile, as a function of time and radius.  Red represents prograde motion, while blue represents retrograde motion.  Black line represents the convective-radiative interface.  Variations are shown in nHz.  Bottom:  Angular velocity variation as a function of time at the convective-radiative interface.  Dotted line represents zero fluctuation about the mean.}
\end{figure}
\clearpage
\begin{figure}
\plotone{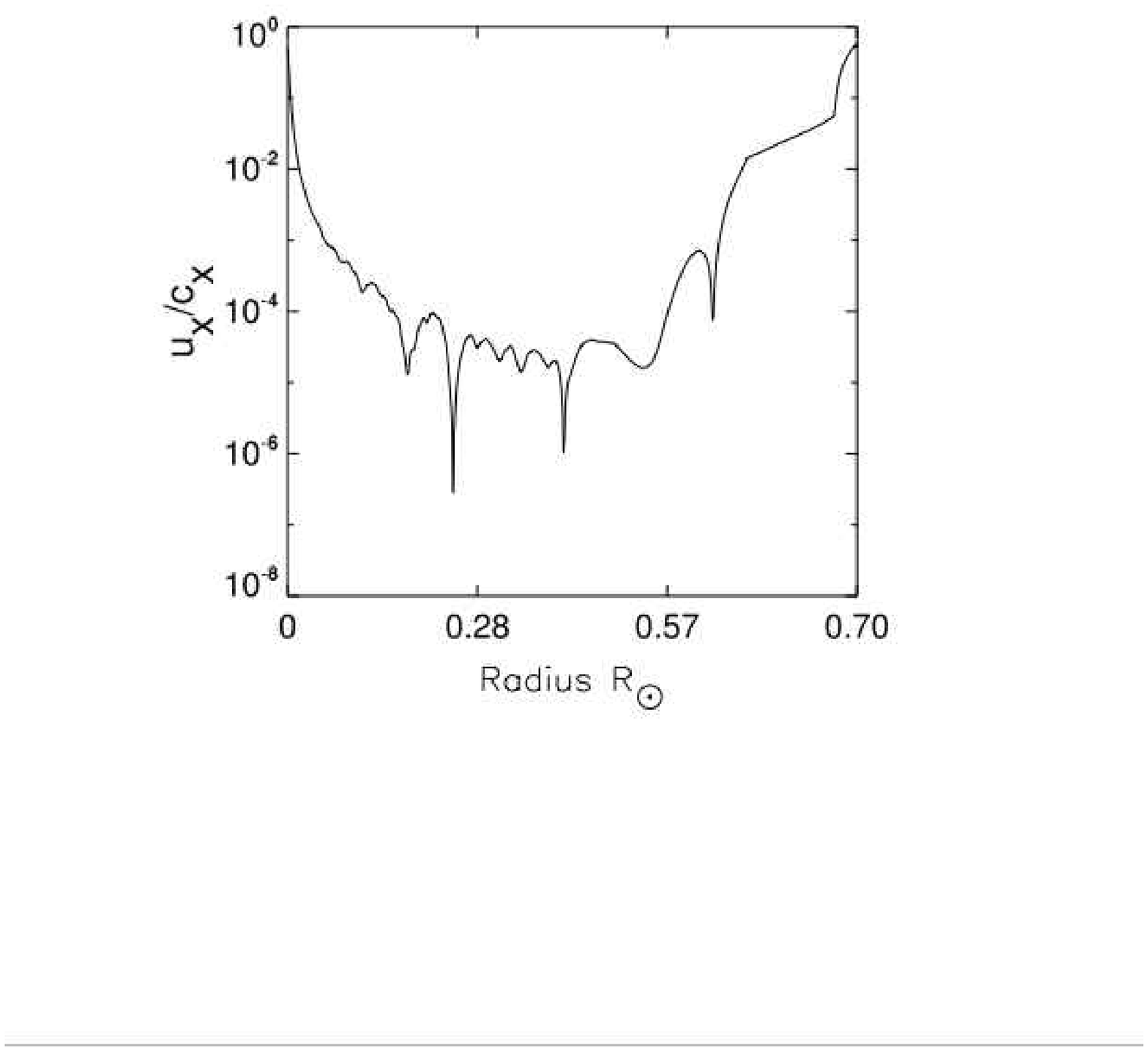}
\caption{The ratio of fluid velocities to the phase velocity (for a typical frequency and wavenumber), also known as the Froude number, a measure of the nonlinearity of the waves.  This figure shows that motions are clearly nonlinear, both in the tachocline and the core.}
\end{figure}
\clearpage
\begin{figure}
\plotone{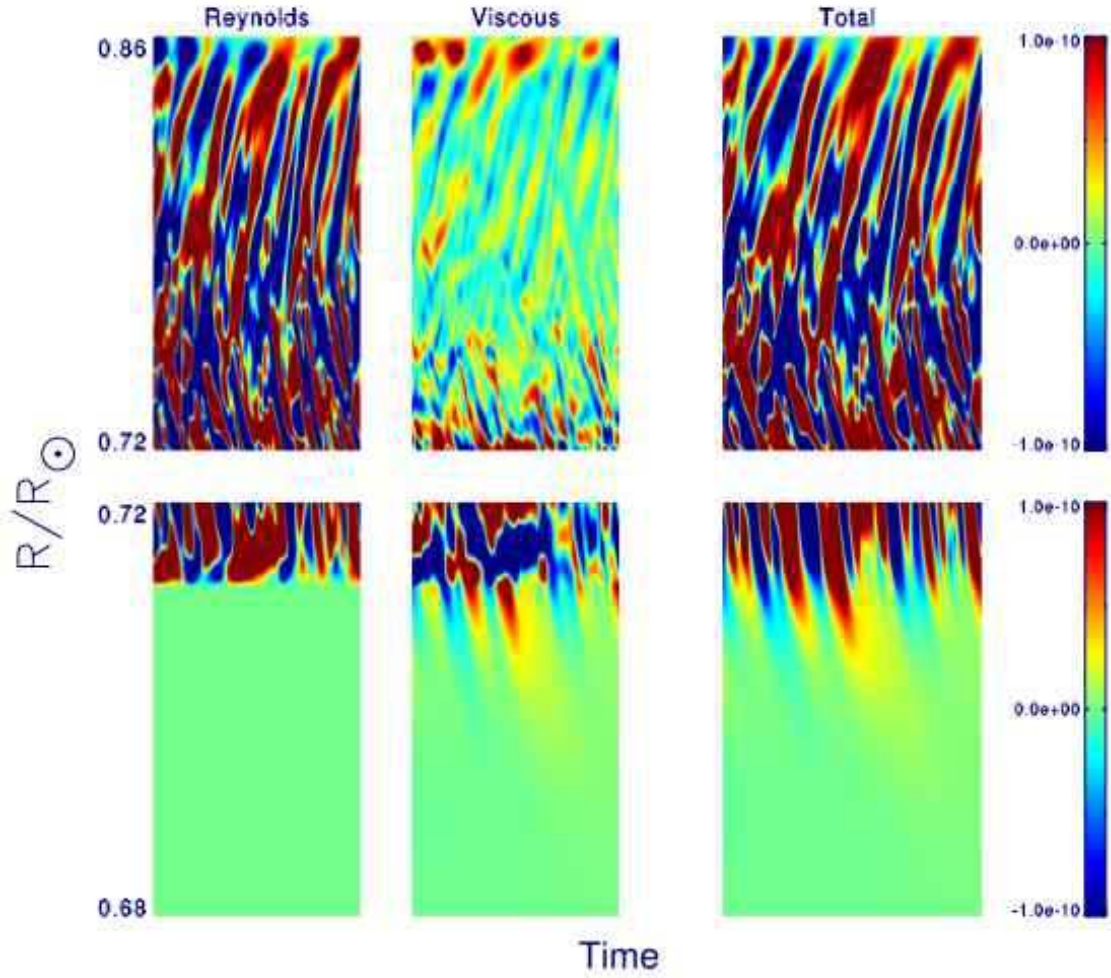}
\caption{Convergence of the Reynolds stress, Viscous stress and the sum, $\partial{\overline{U}}/\partial{t}$, over a five-day period in both the convection zone (top) and overshoot region (bottom).  A semi-periodic oscillation is seen both in the convection zone and overshoot region.  This oscillation causes an oscillation in the amplitude (but not direction) of the mean zonal flow seen in Fig. 1.}
\end{figure}
\clearpage
\begin{figure}
\plotone{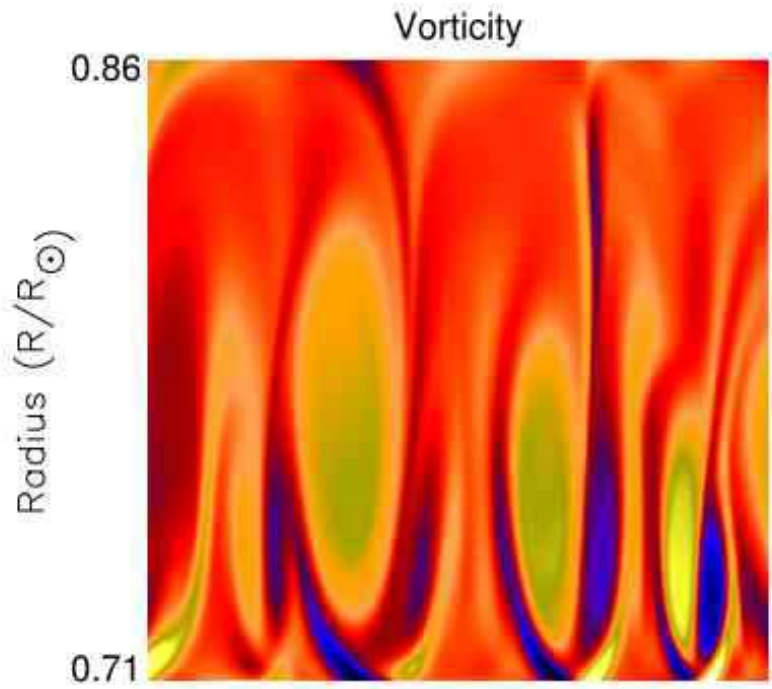}
\caption{Zoom in of vorticity in the convection zone.  Large cells span the bulk of the convection zone with counter-rotating cells beneath due to the deflection of descending plumes as they encounter the stiff radiative region.}
\end{figure}
\clearpage
\begin{figure}
\plotone{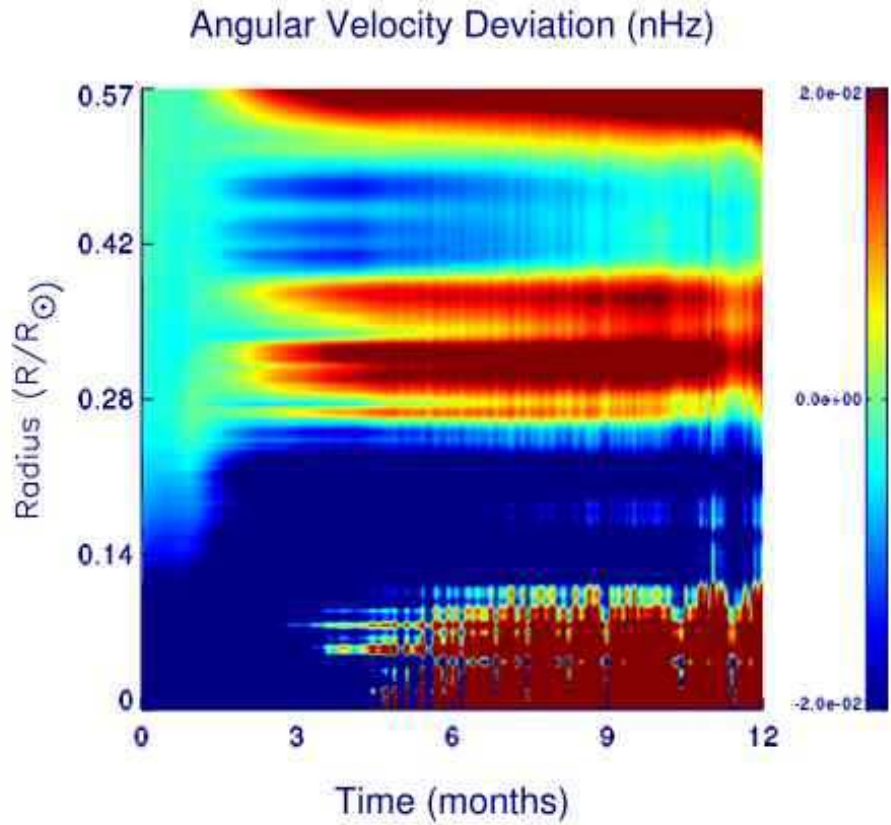}
\caption{Angular Velocity variations in the deep interior.  Banded radial differential rotation is seen.  The deviations in the core are 2-3 orders of magnitude larger than those in the bulk of the radiative interior.  There appears to be a link between deviations in the core and those in the tachocline (see Figure 1).}
\end{figure}
\clearpage
\begin{figure}
\plotone{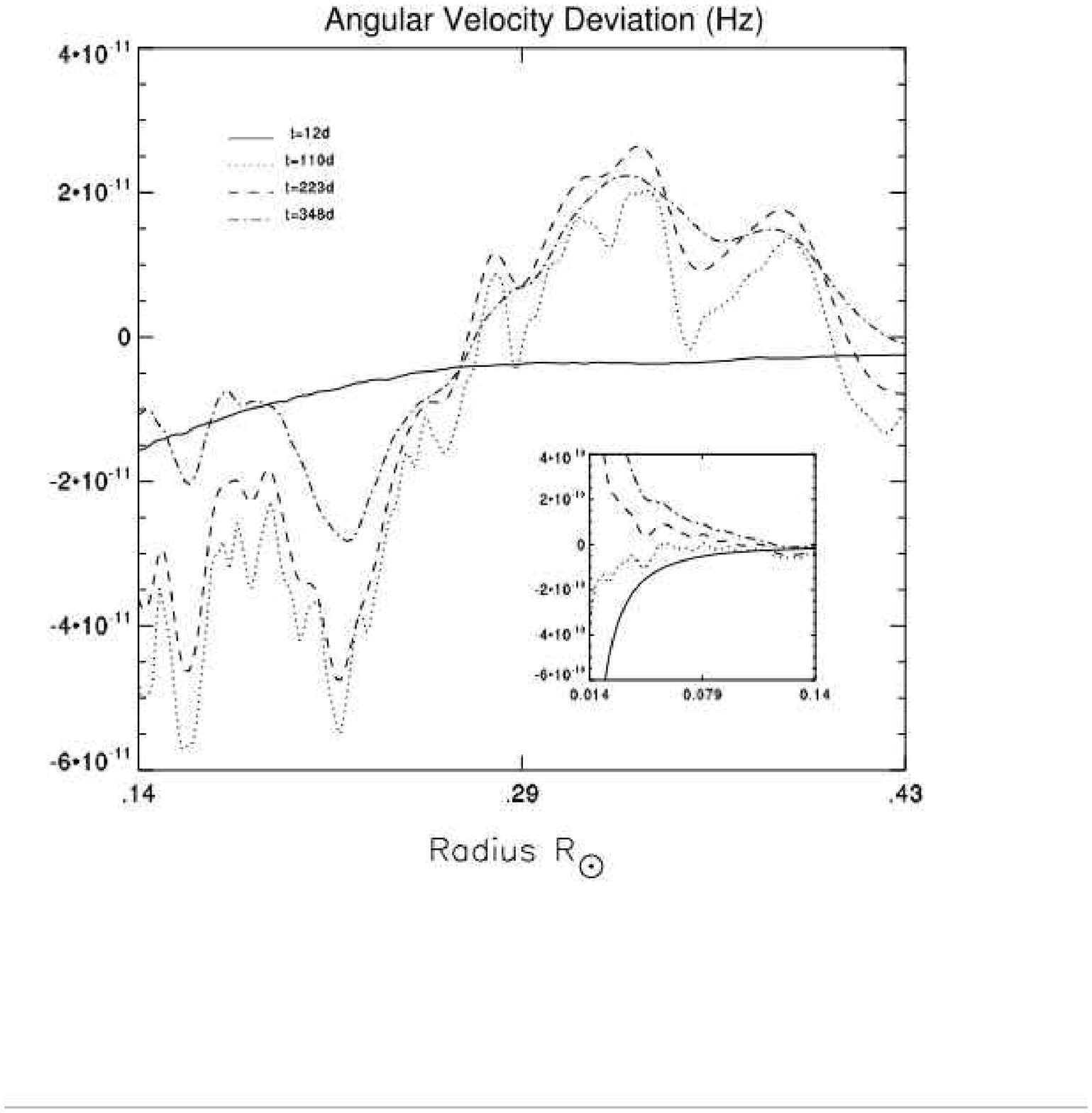}
\caption{Time evolution of angular velocity variations.  Some indication of reversal is seen in the bulk of the radiative interior between 223 days and 348 days.  In the core there was clearly a reversal, which Figure 5 tells us was around 130 days.}
\end{figure}
\clearpage
\begin{figure}
\plotone{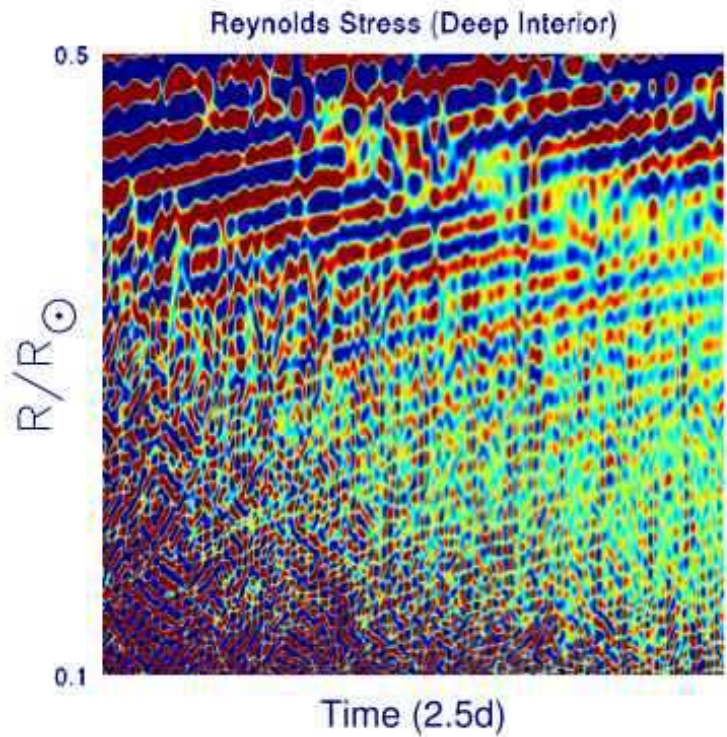}
\caption{Convergence of the Reynolds stress in the bulk of the radiative interior (blue represents negative, red positive values).  Disturbances move upward in time as expected from the nonlinear interaction of low amplitude waves.  Significant interference is seen at smaller radii, probably due to the interaction of inward propagating and reflected waves.}
\end{figure}%


\begin{thebibliography}{}
 \bibitem[Baldwin et al. (2001)]{bal01} Baldwin, M.P., Gray, L.J., Dunkerton, K.Hamilton, Haynes, P.H., Randel, W.J., Holton, J.R. et al. 2001, Rev. Geophysics, 39, 179

\bibitem[Eliassen \& Palm (1961)]{ep61} Eliassen, A. and Palm, E. 1961 Geofysiske Publ., 22, 1
\bibitem[Garcia-Lopez and Spruit (1991)]{gls91} Garcia-Lopez, R.J., Spruit, H. 1991, \apj, 377, 268
\bibitem[Glatzmaier et al. (2005)] Glatzmaier, G.A., Evonuk, M. and Rogers, T.M. 2005, Nature, under review
\bibitem[Gough \& McIntyre (1998)]{gm98} Gough, D.O., McIntyre, M.E. 1998, Nature, 394, 755
\bibitem[Howe et al. (2001)]{how02} Howe, R., Christensen-Dalsgaard, J., Hill, F., Komm, R.W., Larsen, R.M., Schou, J., Thompson, M.J., and Toomre, J. 2001, Science, 287, 2456
\bibitem[Holton (1994)]{hol94]} Holton, J.R., {\it An Introduction to Dynamic Meteorology, 4th edition}, Elsevier Academic Press, 1994
\bibitem[Hurlburt, Toomre \& Massaguer (1986)]{htm86} Hurlburt, N.E., Toomre, J. and Massaguer, J.M. 1986, ApJ, 311, 563
\bibitem[Kim and MacGregor (2001)]{kim01} Kim, E., MacGregor, K.B. 2001, \apj 556, L117
\bibitem[Kumar \& Quataert (1997)]{kq97} Kumar, P., Quataert, E. 1997, ApJ, 475, L143
\bibitem[Kumar, Talon \& Zahn (1999)]{ktz99} Kumar, P., Talon, S., Zahn, J.P. 1999, \apj, 520, 859
\bibitem[Leovy et al. (1991)]{lo91} Leovy,C.B., Friedson, A.J., Orton, G.S. 1991 Nature, 354, 380  
\bibitem[Lindzen (1990)]{lin90} Lindzen, R.S., {\it Dynamics in Atmospheric Physics}, Cambridge University Press, 1990
\bibitem[Plumb and McEwan (1978)]{pmb78} Plumb, R.A., McEwan, A.D. 1978, J. Atmos. Sci, 35, 1827
\bibitem[Press (1981)]{pre81} Press, W.H. 1981, \apj, 245, 286
\bibitem[Ringot (1998)]{ri98} Ringot, O. 1998, A\&A, 335, L89

\bibitem[Rogers \& Glatzmaier (2005)]{rg05} Rogers, T.M., Glatzmaier, G.A. 2005, MNRAS, in press
\bibitem[Schatzman (1993)]{sc93} Schatzman, E. 1993 A\&A, 279, 431 
\bibitem[Talon, Kumar, \& Zahn (2002)]{tkz02} Talon, S., Kumar, P., Zahn, J.P. 2002, \apjl, 574, 175

\bibitem[Talon \& Charbonnel (2005)]{tc05} Talon, S., Charbonnel, C. 2005, \aap, 440, 981
\bibitem[Wedi \& Smolarkiewicz (2005)]{ws05} Wedi, N.P., Smolarkiewicz, P.K., 2005, JAS, submitted
\end{thebibliography}
\end{document}